\documentclass[twocolumn,showpacs,preprintnumbers,prl]{revtex4} 
\usepackage[dvips]{graphicx}
\begin{document}

\title{Energy dependent wavelength of the ion induced nanoscale 
ripple}

\author{T. K. Chini}
\email{tapashp@hp2.saha.ernet.in}
\author{M. K. Sanyal}
\author{S. R. Bhattacharyya}
\affiliation{Surface Physics Division, Saha Institute of Nuclear Physics, 
1/AF Bidhannagar, Kolkata 700 064, India}

\begin{abstract}
Wavelength variation of ion beam induced nanoscale ripple structure has  received much attention recently due to its possible application in nanotechnology. We present here results of Ar$^+$ bombarded 
Si in the energy range 50 to 140 keV to demonstrate that with beam 
scanning the ripple wavelength increases with ion energy and 
decreases with energy for irradiation without 
ion beam scanning. An expression for the energy dependence of ripple 
wavelength is proposed taking into simultaneous effect of thermally 
activated surface diffusion and ion induced effective surface diffusion.
\end{abstract}

\pacs{68.35.Bs, 79.20.Rf, 64.60.Ht} 


\maketitle

Formation of periodic undulations or ripple like features on various 
materials with typical wavelength ranging from about 10 nm to 1$\mu$m, 
obtained by obliquely incident ion bombardment, has become an active  
research subject due to its possible technological applications, 
as varied as optical 
devices, templates for liquid crystal orientation and strain-free patterned 
substrates for heteroepitaxial growth of quantum wires. It is also 
expected that systematic study of ion beam induced nano ripple 
formation will help us to understand the basic processes prevalent 
in formation of sand dune like structures in deserts. 
Although this ion induced phenomenon was reported first in 
1960s \cite{mn1} and then in 1970s \cite{rs2,gc3}, 
the improvement in experimental conditions such as, better vacuum and 
ion beam parameters and improved surface characterizing tools, 
has enabled us to control the growth of these ripple like 
features \cite{je4,gc5,ad6,sh7,ec8,je9,sf10,sr11}. 
The first widely accepted theoretical approach 
describing the process of ripple formation due to ion bombardment 
was developed by Bradley and Harper (BH) \cite{rmb12}. This linear 
theory \cite{rmb12} predicts the ripple wavelength and orientation 
in agreement with numerous experimental studies. However, this theory 
cannot explain a number of 
experimental observations, such as the saturation of the ripple amplitude 
\cite{je9}, the appearance of rotated ripples \cite{sr11} and kinetic 
roughening \cite{ea13}. Moreover, according to the BH theory ripple 
wavelength should decrease with ion energy but this prediction has not 
been confirmed experimentally so far \cite{gc5,ad6,sh7}.

Recently a formalism \cite{mar14,sp15,bk16} based on nonlinear 
continuum theory has been developed to understand these experimental 
observations not predicted by linear theory. In this new formalism, not only 
nonlinear and noise terms were included in the equation of height profile 
for eroded surface but also existence of two different surface diffusion 
processes were recognised. 
Based on Sigmund's theory of sputtering \cite{ps17}, the height 
evolution $h(x,y,t)$ of an ion eroded surface according to this nonlinear 
theory \cite{mar14,sp15,bk16} can be described by
\begin{equation}
\partial_{t}h = \nu\nabla^2h - (D^T + D^I)\nabla^4h + \frac{\lambda}{2}
(\nabla h)^2 + \eta(x,y,t).
\end{equation}
Here $\nu$ is the roughening prefactor (also called effective surface 
tension), generated by the ion bombardment sputter-erosion 
process. $D^T$ \& $D^I$ are designated as thermally activated surface 
diffusion constant and ion induced effective surface diffusion (ESD) constant. 
The nonlinear term $\lambda$ represent the slope dependent erosion rate, 
where $(\nabla h)$ define the local slopes and $\eta(x,y,t)$ is an 
uncorrelated white noise with zero average, mimicking the local random 
fluctuation of the incident ion flux. All the coefficients of Eq. (1) can be 
made directional in $(x,y)$ space and their detailed expressions are known 
\cite{mar14,bk16}. It is interesting to note that beyond a cross-over 
time, $\tau\sim (D/\nu^2)\ln(\nu/\lambda)$, the nonlinear terms become 
dominating and this effect is expected to generate many 
novel phenomena like ripple rotation in oblique ion incidence 
\cite{sr11} and formation of dots and holes for normally incident ion beams 
\cite{bk16}. On the otherhand for time, $t \ll \tau$, nonlinear and noise 
terms can be neglected ($\lambda = 0, \eta = 0$) and Eq. 1 reduces to the 
BH theory \cite{rmb12} except for the fact that surface diffusion constant 
$K$ has two components $K = D^I + D^T$.
In the linear regime the balance of the unstable negative sputter-erosion 
term $(-|\nu|\partial^2{h})$, trying to roughen the surface and the   
positive surface diffusion term $(K\nabla^4h)$ acting to smooth the surface, 
gives rise to observable ripple with wavelength 
\begin{equation}
l_i = 2\pi\sqrt{2K/|\nu_i|},
\end{equation}
where, $i$ refers to the direction $x$ or $y$ along which the associated 
$|\nu_i|$ $(|\nu_x|$ or $|\nu_y|)$ is largest.  

The ion induced ESD process predicts increase of ripple wavelength ($l$) with ion energy ($E$) but the thermal diffusion process, 
assumed to be dominating at high temperature, predicts reduction of $l$ with 
increasing $E$. While the increase of $l$ with $E$ has been observed 
experimentally \cite{ad6,sh7,df18}, direct observation of decreasing $l$ 
with increasing $E$ has not been reported so far to the best of our 
knowledge, even when the sample temperature is varied \cite{gc5}. We have performed 
Ar$^+$ bombardment on Si(001) in 50-140 keV energy region and kept the 
bombardment process well within the linear region to verify experimentally 
the existence of these two diffusion processes and to understand the 
interplay between them in determining the wavelength of nanoscale ripple 
structure.

The ion bombardment of Si targets (p type Si(001) single crystal wafer) 
was undertaken with a high current ion accelerator (Danfysik) which is 
described in detail elsewhere \cite{tc19}.  
These Si samples were irradiated in the implantation chamber 
(having a base vacuum in the range of 10$^{-7}$ mbar) by a 
focused (with typical beam spot size of 1.5 mm and 2.5 mm) $^{40}$Ar$^+$ 
beam at 60$^o$ angle of 
ion incidence with respect to surface normal of the samples. 
With a magnetic $x-y$ sweeping system the focused beam was scanned 
over the sample to maintain homogeneous irradiation. The scanned area 
could be varied from 5 mm $\times$ 5 mm to 15 mm $\times$ 15 mm. 
To check the influence of the beam scanning, several samples were 
irradiated without beam scanning. 
The beam flux in the present experiment was around 140 $\mu$A/cm$^2$ 
for irradiation without beam scanning. 
As the area of the bombarded spot 
on the sample for unscanned focused beam was smaller than that for 
the case of scanned beam, the period of bombardment was adjusted to 
give same dose of 10$^{18}$ ions/cm$^{2}$ for all the irradiated 
samples. The maximum period of bombardment in the present case was always 
less than 1 hr, which is far less than the estimated cross-over time 
$\tau$ (about 3 hrs for 50 keV and 4 hrs for 100 keV) \cite{sp15} ensuring 
the present experimental condition to be well within the linear regime. 
Thus, a higher dose of (3-4)$\times$10$^{18}$ ions/cm$^2$ is required 
to expect any nonlinear effect whose signature has indeed been observed 
in our preliminary results reported elsewhere \cite{dd20} for the case 
of morphology developed on GaAs bombarded by 60 keV Ar beam.

After irradiation, the samples were investigated by atomic force 
microscopy (AFM) in contact mode under ambient condition as 
described earlier \cite{tc19,dd20,jb21}.
Fig.~1 shows representative AFM images of Si(001) samples bombarded 
at 50 and 100 keV with and without beam scanning. Ripple 
topographies are clearly visible in all the AFM micrographs and 
they are oriented perpendicular to the ion-beam projection (indicated 
by arrowmarks) onto the surface as predicted by linear theory. 
The ripples grew uniformly on the whole bombarded 
spot for irradiation with beam scanning whereas ripples were 
developed at the central region of the bombarded spot for 
irradiation without scanning.  
The mean amplitude of the ripples with and without beam scanning, 
obtained from AFM measurements, are also shown in fig.~1.
The most probable observed wavelength ($l$) of ripples, obtained from AFM 
images and Fourier analysis is given in fig.~2. With beam scanning 
$l$ grows from about 700 nm to about 1000 nm 
as the incident ion energy (E) increases from 50 keV to 140 keV.
However, the $l$ decays from 600 nm to about 400 nm 
with the increasing $E$ when the ripple is formed  
without beam scanning. If thermally activated surface 
diffusion is the dominant process for surface smoothing, the wavelength 
dependence on energy ($E$) of bombardment is given by \cite{rmb12} 
$l \sim E^{-1/2}$.
On the otherhand in the absence of thermal diffusion, ion-induced 
ESD becomes dominating contributor in $K$ and we expect \cite{mar14} 
the energy dependence as $l \sim E$.
The variation of $l$ with $E$ of the present 
experiment has exhibited both the trends : one increasing and 
other decreasing with the 
energy giving rise to exponents of 0.45 and -0.56 respectively (refer Fig.~2). 

When the sample is bombarded with a focused beam without scanning,  
the beam intensity at the center of the bombarded spot becomes more 
than its peripheral region causing a local temperature rise.
As the semiconductor sample in the present case is clamped on a 
copper block, good thermal contact is not ensured favouring the local 
temperature rise due to ion beam heating. Under this situation, 
the rise of temperature, $T$, from the initial ambient temperature , 
$T_o = 25^oC$, at time t(sec) of the commencement of the bombardment can 
be calculated from the formula \cite{gd22}
\begin{equation}
t = \int_{T_o}^{T} \frac{S\kappa\rho}{P - (T^4 - T_o^4)}dT,
\end{equation}
where $S$,$\kappa$ and $\rho$ are specific heat (0.836 $J/g/^oC$), 
thickness (0.47 mm) and density (2.32 $g/cm^3$) respectively  
of the Si wafer used in the present experiment.
Sample temperature thus estimated within 68 sec \cite{gd22} after the start of 
bombardment was $\sim$800$^oC$ for a beam power, P = 
14 watt/cm$^2$ corresponding to the maximum flux (140$\mu$A/cm$^2$) 
employed in the present case for bombardment at 100 keV.
As the period of bombardment is much higher than 
this rise time, beam induced heating is expected to be appreciable 
in the present case for the bombardment without scanning. 
Thus, thermally activated surface diffusion plays an important role 
here giving rise to a negative exponent $l\sim E^{-0.56}$, as expected 
from linear theory. It seems, the surface temperature profile on the 
sample generated by unscanned focused ion beam has made thermally 
activated diffusion process the dominating one. This was not the case even 
when experiments were performed by increasing the bulk temperature 
of the sample \cite{gc5}. It should be mentioned here that 
we did not observe any flux dependence of the measured $l$ 
in the flux range of 15 to 140 $\mu$A/cm$^{2}$  
used here as predicted in BH model.  

On the otherhand, for irradiation with beam scanning, effective beam 
intensity on the bombarded spot is reduced as the beam is 
sweeped continuously over the sample. The magnetic beam sweeping system 
of our implanter uses scan frequencies of 0.51 Hz and 5.3 Hz along the 
horizontal and vertical direction to maintain beam homogeneity on the 
irradiated area of the sample. This means, each region of the 
bombarded spot of the size of beam diameter will be exposed to the 
ion beam every after few seconds only which is much less than 
the time of rise of the maximum equilibrium temperature due to beam 
heating. So promotion of surface diffusion of thermal origin will be 
hampered and ion induced 
surface diffusion will be the main relaxation process in the case of 
beam scanning. However, at relatively lower energy thermal contribution 
may become noticeable.

We can rewrite the ripple wavelength expressed by Eq.~2 
by splitting the two diffusion terms as 
\begin{equation}
l^2 = 8\pi^2\frac{D^T + D^I}{|\nu|}
\end{equation}
which can be approximated as 
\begin{equation}
l^2 = A + BE^{-1} + CE^2
\end{equation}
where A, B and C are fitting parameters. 
In this approximation, we have assumed that both depth ($a$) 
and longitudinal spread ($\alpha$) are linear with energy, as 
obtained exponents from TRIM calculations \cite{tr23} are close to unity 
(refer Fig.~2). 
It is possible to 
determine the shape of the distribution of the deposited energy density 
in the near surface region induced by the ion collision cascades. As for 
example, for the case of ripple topography formation on Ar bombarded 
graphite in 2 -50 keV energy region Habenicht \textit{et al} \cite{sh7} 
obtained 
a power law relation $l(E) \sim E^p$ ($p = 0.95$) while $a$ 
and $\alpha$ scale with ion energy implying an 
energy-independent lateral spread of the damage cascade. However, 
we obtained $l(E) \sim E^p$ ($p = 0.45$) with beam scanning (Fig.~2). 
A lower exponent ($p \ll 1$) 
is an indication of the changing lateral spread of the damage cascade 
with the ion energy for the energy region selected in the present 
experimental situation.  
The modified expression (Eq.~5) of the energy 
dependence of ripple wavelength now considers the simultaneous 
effect of ion induced effective surface diffusion and thermally 
activated surface diffusion.

The growth and decay of $l$ with $E$ of the present experimental 
results (Fig.~3a) can be fitted with Eq. 5 with fitting 
parameters $A = 311697$, $B = 2.5 \times 10^6$ and $C = 41.5$ for the ripple 
formation with beam 
scanning. The corresponding values for the ripple wavelength without 
beam scanning is $A = -195533$, $B = 2.9 \times 10^7$ and $C = 8.0$. 
Here $l$ and $E$ are expressed in nm and keV respectively. 
The higher value of the coefficient of the second term for fitting 
the ripple wavelength data without beam scanning 
signifies that contribution of thermally activated surface diffusion 
is higher than that prevalent with beam scanning.
 
To test the validity of our proposed 
formalism we have also used it for the data of Flamm \textit{et al} \cite{df18} 
where ripple structures with a characteristic wavelength between 30 and 
300 nm are formed on fused silica at room temperature after irradiation 
by broad beam ion (Ar) source 
in the energy region 0.6 to 1.5 keV. The expected dominance of ion 
induced effective surface diffusion has been reflected in this case 
by the excellent fit of the data (Fig.~3b)  
with Eq. 5 having much higher value 
of the coefficient of the ion induced diffusion term ($C = 19861$) 
over the value 
of the coefficient ($B = 6077.8$) representing the thermal diffusion term.
Thus, our proposed formula behaves consistently with homogeneous 
bombardment condition from the low ($\sim$ 1 keV) to the 
medium ($\sim$ few hundreds keV) energy regime where the concept of linear 
collision cascade model holds for heavy ion bombardment on solid target. 

In summary, we have experimentally verified, for the first time, the 
existence and interplay of two types of diffusion mechanism prevalent 
in ion induced nanoscale nanoscale ripple formation giving rise to 
the increasing or decreasing tendency of the ripple wavelength 
with ion energy.
We have proposed a formalism for the ripple wavelength variation 
with the ion energy and verified its applicability over a wide energy range.   
The proposed formalism indicates that both thermal and ion induced 
surface diffusion act simultaneously to form stable ripple morphology.  
Ion beam induced surface temperature profile, rather than bulk 
temperature of the sample, seems to initiate the thermally activated 
diffusion process. It will be interesting to investigate the role of 
this temperature profile in the formation of uniform nanoscale ripple 
structure and in the nonlinear phenomena.

We would like to thank Mr. S. Roy and Mr. A. Das for their 
technical assistance and Dr. S. Hazra for valuable discussions. 

\newpage
\begin{figure}[hbp]
\caption{
\label{fig1}AFM micrographs of $^{40}$Ar$^+$ bombarded Si 
surfaces at two representative 
energy, namely, 50 keV and 100 keV : (a) and (b) with ion beam scanning; 
(c) and (d) without beam scanning. Arrow marks indicate the 
projection of ion beam onto the surface. Variation of 
mean amplitude of the ripples with ion energy is shown in the lower plot.}
\end{figure}
\begin{figure}[hbp]
\caption{
\label{fig2}Left scale depicts the variation 
of the mesaured wavelengths $l$ as a function of the $^{40}$Ar$^+$ energy 
and the right scale shows the TRIM calculations for depth $a$ and 
longitudinal straggling $\alpha$ as a function of ion energy.
Power law fits to the wavelengths and TRIM data are inserted to each 
plot.
The error bar of the measured data points have been determined after 
several repeated experiments and each data point represents the averages 
over different regions of the bombarded spot.}
\end{figure} 
\begin{figure}[hbp]
\caption{\label{fig3}
(a) The square of the measured ripple wavelengths of the present experiment 
are fitted with Eq. (4). The function 
with the fitted parameters for the ripple wavelength data 
obtained with and without beam scanning are inserted within the graph.
(b) The fitting of the data extracted from the measurements of Flamm 
\textit{et al} \cite{df18}}
\end{figure}
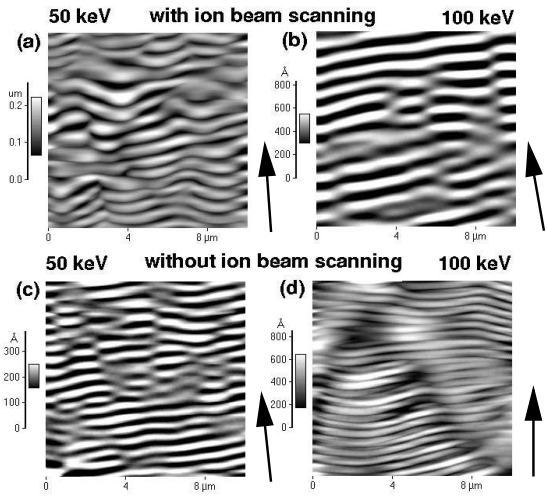
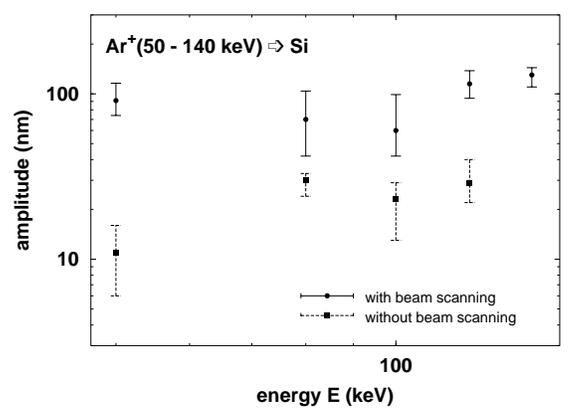
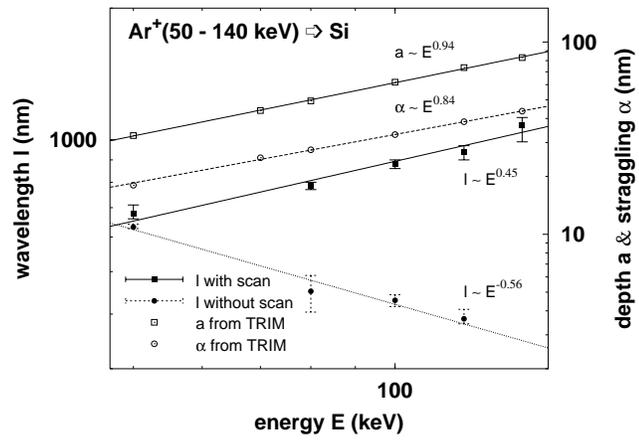
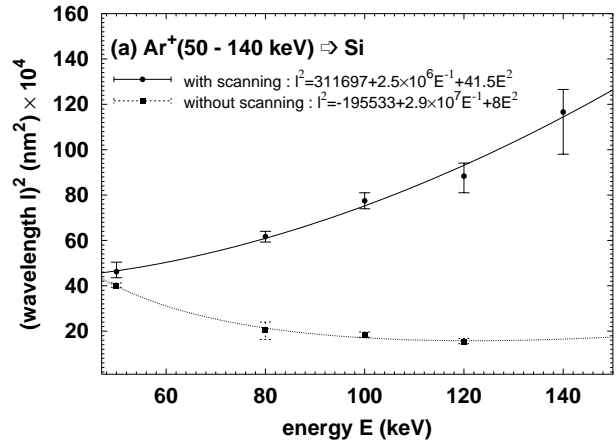
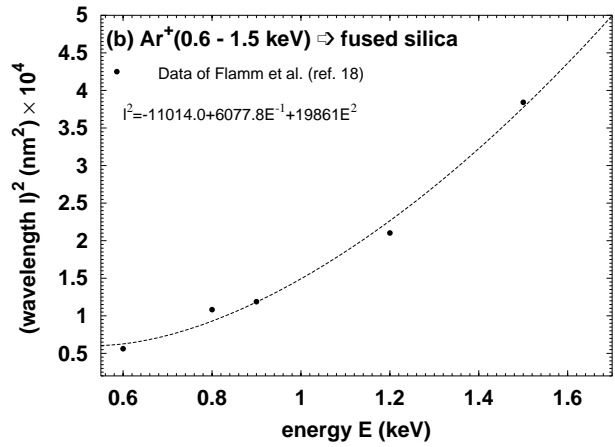
\end{document}